\documentclass[aps,prd,preprint,groupedaddress]{revtex4-1}
\usepackage[dvips]{graphicx}
\usepackage{color}

\def\nn{\nonumber}
\def\eq{equation}
\def\eqn{eqnarray}

\begin{document}

\title{Fourth-order split monopole perturbation solutions \\
 to the Blandford-Znajek mechanism}
\author{Zhen Pan}
\email{zhpan@ucdavis.edu}
\affiliation{Department of physics, University of California, Davis}
\author{Cong Yu}
\email{cyu@ynao.ac.cn} \affiliation{Yunnan Observatories, Chinese
Academy of Sciences, Kunming, 650011, China \\
Key Laboratory for the Structure and Evolution of Celestial
Object, Chinese Academy of Sciences, Kunming, 650011, China}
\date{\today}

\begin{abstract}
The Blandford-Znajek (BZ) mechanism describes a physical process
for the energy extraction from a spinning black hole (BH), which
is believed to power a great variety of astrophysical sources,
such as active galactic nuclei (AGNs) and Gamma ray bursts (GRBs).
The only known analytic solution to the BZ mechanism is a split
monopole perturbation solution up to $O(a^2)$, where $a$ is the
spin parameter of a Kerr black hole. In this paper, we extend the
monopole solution to higher order $\sim O(a^4)$. We carefully
investigate the structure of the BH magnetosphere, including the
angular velocity of magnetic field lines $\Omega$, the toroidal
magnetic field $B^\phi$ as well as the poloidal electric current
$I$. In addition, the relevant energy extraction rate $\dot E$ and
the stability of this high-order  monopole perturbation solution
are also examined.
\end{abstract}


\maketitle

\section{Introduction}
Within the framework of force-free electrodynamics, Blandford \&
Znajek (1977) investigated a steady-state axisymmetric
magnetsphere surrounding a spinning black hole and put forward
that the rotation energy of a Kerr black hole could be extracted
in the form of Poynting flux via magnetic fields penetrating the central
black hole (\cite{Blandford1977d,2000PhR...325...83L}). General
relativistic magnetodynamics (GRMD) simulations of split monopole
magnetic fields (\cite{Komissarov2001,Komissarov2004e}) show that
the analytic monopole perturbation solution makes good matches
with the numerical simulations, especially for slowly rotating
black holes. General relativistic magnetohydrodynamics (GRMHD)
simulations (\cite{McKinney2004f,2005ApJ...630L...5M}) indicate
that, in the polar region, the monopole perturbation solution
gives a good description of the magnetic field configuration as
well as the angular distribution of energy flow, even when black
holes rotate mildly rapidly. However, the monople solution
\cite{Blandford1977d} is accurate
only up to $O(a^2)$, where $a$ is the black hole spin parameter.
For even more rapidly rotating black holes, higher order
perturbation solutions are of greater astrophysical interests.
\citet{Tanabe2008} extended the  monopole perturbation
solution to the order of $O(a^4)$. Their solution gave a better
approximation to the numerical simulation. Unfortunately, they
mentioned that their results are not fully self-consistent, since
their perturbation method breaks down at large distance from the
central black hole. Hence, it is necessary to find self-consistent
higher order perturbation solutions to the BZ mechanism.

To get self-consistent solutions, we need to solve a nonlinear
second-order partial differential equation, which requires two
boundary conditions. It should be noted that boundary conditions
to be imposed are still not well understood
(\cite{2004ApJ...603..652U,2005ApJ...620..889U,Tanabe2008,2004astro.ph..9076B,2014ApJ...788..186N}).
\citet{Blandford1977d} imposed the Znajek regularity condition
(\cite{Znajek1977b}) on the horizon as the first boundary
condition. The second one requires that the perturbation solution
should match the asymptotic solution in the flat spacetime at
infinity (\cite{1973ApJ...180L.133M}). Unfortunately, the second
boundary condition is usually unavailable when investigating
higher order perturbation solutions.
Recently, \citet{Pan2014} proposed that the physical constraint,
i.e., solutions should be convergent from the horizon to infinity,
could be exploited as the second boundary condition. With the
Znajek horizon regularity condition and this new convergence
constraint, perturbation solutions could be uniquely determined.
Following the approach of \citet{Pan2014}, we extend the monopole
perturbation solution to the order of $O(a^4)$. Note that the
perturbation method we adopt is different from \cite{Tanabe2008}.
Our method works well at any distance from the central black hole.

Some earlier analytic works
(\cite{Li2000,Lyutikov2006e,Giannios2006}) concerned the stability
of jets launched by the BZ mechanism because of the screw
instability of the magnetic field. However, such instability was
not found in recent simulations (e.g. \cite{McKinney2009a,
Tchekhovskoy2010, McKinney2013}). The possible reason for the
discrepancy is that the Krustkal-Shafranov (KS) criteria is used
in these works, without taking account of the stabilizing effect
induced by the magnetic field rotation
(\cite{Tomimatsu2001,McKinney2009a,O'Neill2012}). With the
high-order perturbation solution obtained in this paper, we also
briefly study the stability of the split monopole perturbation
solution of the order of $O(a^4)$, taking the magnetic field
rotation into consideration.

The paper is organized as follows: basic equations governing
stationary axisymmetric force-free fields around Kerr black holes
are described in section II. We discuss the perturbation solutions
of second-order and fourth-order obtained by our newly proposed
method in section III. Summary and discussion are given in section
IV.

\section{Stationary Axisymmetric Force-Free Fields around Kerr Black Holes}

In this section, we briefly recap basic equations governing
stationary axisymmetric force-free fields around Kerr black holes
(see \cite{Pan2014} and references therein for more details). We
adopt the Kerr-Schild coordinate (e.g., \citet{McKinney2004f})
with the line element
\[
ds^2 = -\left( 1-\frac{2r}{\Sigma} \right)dt^2 + \left( \frac{4
r}{\Sigma} \right) dr dt + \left(1+\frac{2r}{\Sigma} \right) dr^2
+ \Sigma d\theta^2 - \frac{4 a r \sin^2\theta}{\Sigma} d\phi dt
\]

\begin{\eq}
- 2 a \left(1+\frac{2r}{\Sigma}\right) \sin^2\theta d\phi dr  +
\sin^2\theta \left[\Delta  + \frac{2r (r^2 + a^2) } {\Sigma}
\right] d\phi^2 \ ,
\end{\eq}
where $\Sigma=r^2+a^2\cos^2\theta$, $\Delta=r^2-2r+a^2$, and
$\sqrt{-g}=\Sigma\sin\theta$ .

The energy momentum tensor for the force-free field is dominated
by the electro-magnetic field, which can be written as $T^{\mu\nu}
= T^{\mu\nu}_{\rm matter} + T^{\mu\nu}_{\rm EM}\approx
T^{\mu\nu}_{\rm EM} = F^{\mu\tau} F^{\nu}_{\phantom{nu}\tau} -
\frac{1}{4} \delta^{\mu\nu} F^{\alpha\beta} F_{\alpha\beta}$,
where the  antisymmetric Faraday tensor is defined as $F_{\mu\nu}
= \partial_\mu A_\nu -
\partial_\nu A_\mu$ and $A$ is the $4-$potential of
electromagnetic field. We define the angular velocity of the
magnetic field $\Omega(r,\theta)$ as follows,
\begin{\eq}
- \Omega \equiv \frac{d A_t}{d A_{\phi}} =
\frac{A_{t,\theta}}{A_{\phi,\theta}} = \frac{A_{t,r}}{A_{\phi,r}}
\ .
\end{\eq}
It is evident that $F_{t\phi} = 0$ for the axisymmetric and steady
state force-free field. The non-zero parts of the Faraday tensor
$F_{\mu\nu}$ are listed below:
\begin{\eq}\label{poloidal}
F_{r\phi} = -F_{\phi r}=A_{\phi,r} \ , F_{\theta\phi} = -
F_{\phi\theta} = A_{\phi,\theta} \ ,
\end{\eq}
\begin{\eq}\label{electricfield}
F_{tr} = -F_{rt} = \Omega A_{\phi,r} \ , F_{t\theta}= - F_{\theta
t} = \Omega A_{\phi,\theta} \ ,
\end{\eq}
\begin{\eq}\label{toroidal}
F_{r\theta} = -F_{\theta r} = \sqrt{-g}B^{\phi} \ .
\end{\eq}
Note that the force-free field is specified by three quantities,
i.e., $\Omega(r,\theta)$, $A_{\phi}(r,\theta)$, and
$B^{\phi}(r,\theta)$. Once they are specified, the force-free
field is uniquely determined.

Note that $T^{\theta}_{\phantom{\theta}t} = -\Omega
T^{\theta}_{\phantom{\theta}\phi}$ and   $T^{r}_{\phantom{r}t} =
-\Omega T^{r}_{\phantom{r}\phi}$.    %  \ .
The energy and angular momentum conservation equations
$T^{\mu}_{\phantom{\mu}t;\mu} = 0$ and
$T^{\mu}_{\phantom{\mu}\phi;\mu} = 0$ can be cast as
$\Omega_{,r}A_{\phi,\theta} = \Omega_{,\theta}A_{\phi,r}$ and
$(\sqrt{-g}F^{\theta r})_{,r}A_{\phi,\theta} = (\sqrt{-g}F^{\theta
r})_{,\theta}A_{\phi,r}$, respectively. These two equations
indicate that $\Omega$ and $\sqrt{-g} F^{\theta r}$ are functions
of $A_\phi$, viz,
\begin{\eq}\label{constraint}
\Omega \equiv \Omega(A_\phi) \ , \ \sqrt{-g}F^{\theta r} \equiv
I(A_\phi) \ ,
\end{\eq}
where the angular velocity of magnetic field $\Omega$ and the
poloidal electric current $I$ are to be specified. Substituting
Equations (\ref{poloidal}), (\ref{electricfield}),
(\ref{toroidal}) and (\ref{constraint}) into the equation
$F^{\theta r} = g^{\theta\mu} g^{r\nu} F_{\mu\nu}$, we can readily
arrive at
\begin{\eq}\label{znajek}
B^\phi = - \frac{I \Sigma + (2 \Omega r - a) \sin\theta
A_{\phi,\theta}} {\Delta \Sigma \sin^2\theta} \ .
\end{\eq}
This is an important relation that connects the toroidal magnetic
field $B^{\phi}$ with the functions $A_{\phi}(r,\theta)$,
$\Omega(A_{\phi})$ and $I(A_{\phi})$.

The remaining momentum conservation equations in the $r$ and
$\theta$ direction $T^{\mu}_{\phantom{\mu}r;\mu} = 0$ and
$T^{\mu}_{\phantom{\mu}\theta;\mu} = 0$ are actually equivalent
and read
\begin{\eq}\label{grad-shafranov}
-\Omega \left[(\sqrt{-g}F^{tr})_{,r} +
(\sqrt{-g}F^{t\theta})_{,\theta} \right] + F_{r\theta}I'(A_\phi) +
\left[(\sqrt{-g}F^{\phi r})_{,r} +
(\sqrt{-g}F^{\phi\theta})_{,\theta} \right] = 0 \ ,
\end{\eq}
where the prime denotes derivative with respect to $A_{\phi}$. The
three functions $A_{\phi}(r,\theta)$, $\Omega(A_{\phi})$, and
$I(A_{\phi})$ are related by the above nonlinear equation
(\ref{grad-shafranov}), which is also widely known as the
Grad-Shafranov (GS) equation
(\cite{2005ApJ...620..889U,2013ApJ...765..113C}).

\section{Fourth-order Perturbation Solutions}
{\bf Since the Farady tensor depends on the first order derivative of
$A_{\phi}$, it is clear that the GS equation
(\ref{grad-shafranov}) is actually a second-order partial
differential equation for $A_\phi$.} The solution can be attained
when complemented with two boundary conditions, i.e., the Znajek
horizon regularity condition (\cite{Znajek1977b}) and the
convergence constraint (\cite{Blandford1977d, Pan2014}). The
zeroth-order monopole solution can be readily obtained when the
black hole is non-rotating, i.e., $a=0$. When the spin parameter
$a\neq0$, we expand the GS equation in terms of $a$. To get the
second-order perturbation solutions, we ignore all terms in the GS
equation that are higher than the order of $O(a^2)$. Based on the
second-order solutions, the fourth-order perturbation solution can
be achieved in a similar way.

The zeroth-order monopole solution around non-rotating black hole
can be explicitly written as (\cite{Blandford1977d}),
\begin{\eqn}
\Omega_0 =0 , \qquad B^{\phi}_0 = 0 , \qquad A_{\phi} = A_0 =
-\cos\theta.
\end{\eqn}
Since $\Omega$ and $\sqrt{-g}F^{\theta r}$ are functions of
$A_{\phi}$, we can expand them, accurate to the order of $O(a^4)$,
as
\begin{\eqn}
\Omega&=&\Omega(A_\phi) = a\omega_1(A_\phi) + a^3\omega_3(A_\phi) =a\omega_1(A_0+a^2A_2)+a^3\omega_3(A_0+a^2A_2),\nn\\
\sqrt{-g}F^{\theta r}&=& I(A_\phi)= ai_1(A_\phi)+a^3i_3(A_\phi)=ai_1(A_0+a^2A_2)+a^3i_3(A_0+a^2A_2),
\end{\eqn}
where $\Omega$, $\omega_1$, $\omega_3$, $I$, $i_1$, $i_3$ are
unknown functions of $A_\phi$ to be specified self-consistently.
The entire fourth-order perturbation solutions can be expressed in
a more compact form as,
\begin{\eqn}
A_{\phi}&=&A_0+a^2A_2+a^4A_4+O(a^6) \ ,\nn\\
\Omega  &=&a\Omega_1+a^3\Omega_3+O(a^5) \ ,\nn\\
\sqrt{-g}F^{\theta r}&=& a I_1 + a^3 I_3+O(a^5) \ , \nn\\
B^{\phi}&=&a B_1+a^3 B_3+O(a^5) \ .
\end{\eqn}
It should be noted that $\Omega_n$ and $\omega_n$, $I_n$ and $i_n$
($n=1,3$) are related by
\begin{\eqn}\label{notation}
\Omega_1&=& \omega_1(A_0) \ , \qquad \Omega_3=\omega_1'(A_0)A_2+\omega_3(A_0) \ , \nn\\
I_1&=&i_1(A_0) \ , \qquad \,  I_3=i_1'(A_0)A_2+i_3(A_0) \ ,
\end{\eqn}
where the prime designates the derivative with respect to $A_0$.

\subsection{Second-order Perturbation Solutions}
We can get the second-order perturbation solutions by expanding
the GS equation (\ref{grad-shafranov}) to the order of $O(a^2)$.
It is interesting that the original BZ monopole perturbation
solution could be naturally achieved with our convergence
constraint. Expanding  Eq.(\ref{znajek}) to the order of $O(a^2)$,
we have that
\begin{\eqn}
r^2I_1 = \sin\theta A_{0,\theta}(1-2r\Omega_1)-\sin^2\theta B_1(r^2-2r)r^2.
\end{\eqn}
According to the Znajek horizon condition (\cite{Znajek1977b}),
the toroidal field $B_1$ should be well-behaved on the horizon
$(r=2)$, then $r=2$ must be a root to equation
$r^2I_1=\sin^2\theta(1-2r\Omega_1)$. So we have
\begin{\eqn}\label{h1w1}
i_1 &=& I_1 = \sin^2\theta \left(\frac{1}{4}-\Omega_1\right),\nn\\
B_1 &=& -\frac{1}{r^2} \left(\frac{1}{4}-\Omega_1+\frac{1}{2r}\right).
\end{\eqn}
The GS equation (\ref{grad-shafranov}), accurate to the order of
$O(a^2)$, can then be cast as
\begin{\eq}
\mathcal L A_2 = S(r,\theta) \ ,
\end{\eq}
where the operator
\begin{\eqn}
\mathcal L\equiv\frac{1}{\sin\theta}\frac{\partial}{\partial
r}\left(1-\frac{2}{r}\right)\frac{\partial}{\partial r}
 +\frac{1}{r^2}\frac{\partial}{\partial \theta}\frac{1}{\sin \theta}\frac{\partial}{\partial \theta} \ ,
\end{\eqn}
and the source
\begin{\eqn}
S(r,\theta) &=&
4\sin\theta\cos\theta\left(\Omega_1-\frac{1}{8}\right)\left(\frac{1}{4}+\frac{1}{2r}+\frac{1}{r^2}\right)
-2\sin\theta\cos\theta\frac{1}{r^2}\left(\frac{1}{2r}+\frac{1}{r^2}\right)\nn\\
& & +\sin^2\theta \Omega_{1,\theta}\left(\frac{1}{4}+\frac{1}{2r}+\frac{1}{r^2}\right).
\end{\eqn}
According to \citet{Blandford1977d}, the condition for the
existence of solution is that the following integral,
\begin{\eqn}\label{convergence}
\int_2^\infty dr \int_0^\pi d\theta \frac{|S(r,\theta)|}{r} \ ,
\end{\eqn}
should be convergent. The convergence condition requires all the
terms in $S(r,\theta)$ of the order of $O(1)$ should vanish, i.e.,
\begin{\eqn}
0 = 4\sin\theta\cos\theta\left(\Omega_1 - \frac{1}{8}\right) +
\sin^2\theta\Omega_{1,\theta} \ .
\end{\eqn}
Consequently, we have
\begin{\eqn}\label{2nd}
\Omega_1&=&\omega_1=\frac{1}{8} \ ,\nn\\
I_1 &=& i_1 = \frac{1}{8}\sin^2\theta \ ,\nn\\
B_1 &=& -\frac{1}{r^2}\left(\frac{1}{8}+\frac{1}{2r}\right) \ .
\end{\eqn}
It is interesting to note that all physical quantities of the
order $O(a)$ are already obtained before we actually solve the
complicated GS equation. The second-order part of $A_{\phi}$,
i.e., $A_2$, can be obtained by the following equation,
\begin{\eq}
\mathcal L A_2 = -2 \sin\theta \cos\theta \frac{1}{r^2}
\left(\frac{1}{2r} + \frac{1}{r^2}\right) \ .
\end{\eq}
It is straightforward while tedious to check that this equation
has the following variable separable
solution(\cite{Blandford1977d})
\begin{\eq}
A_2 = R(r) \sin^2\theta\cos\theta,
\end{\eq}
where
\[
R(r) = \frac{1+3r-6r^2}{12} \ln\left(\frac{r}{2}\right) +
\frac{11}{72}+\frac{1}{3r} + \frac{r}{2}-\frac{r^2}{2}
\]
\begin{\eq}
\qquad\qquad\qquad\qquad + \left[\mathrm{Li}_2\left(\frac{2}{r}\right) -
\ln\left(1-\frac{2}{r}\right)\ln\left(\frac{r}{2}\right)
\right]\frac{r^2(2r-3)}{8} \ ,
\end{\eq}
and
\begin{\eq}
\mathrm{Li}_2(x) = - \int_0^1 dt \frac{\ln(1-tx)}{t} \ .
\end{\eq}
The value of the function $R(r)$ at the horizon, $r=2$, is of
particular importance. Explicitly, it is
\begin{\eq}
R_{r=2} = R({r=2})  = \frac{6\pi^2-49}{72} \ .
\end{\eq}

\subsection{Fourth-order Perturbation Solutions}
Once the second-order perturbation solutions are known, the
fourth-order perturbation solutions could be obtained by further
expanding the GS equation to the order of $O(a^4)$ . Accurate to
$O(a^4)$, Eq.(\ref{znajek}) is
\[
r^2I_3+\cos^2\theta I_1 = \sin\theta
\left[A_{2,\theta}(1-2r\Omega_1) - A_{0,\theta}2r\Omega_3\right]
\]
\begin{\eq}
\qquad\qquad\qquad - \sin^2\theta \left[B_3(r^2-2r)r^2 +
B_1(r^2-2r)\cos^2\theta +B_1 r^2 \right] \ .
\end{\eq}
The toroidal field $B_3$ should be well behaved on the horizon.
Subsequently, we can get
\begin{\eqn}\label{h3w3}
i_3 +\sin^2\theta\omega_3 = \left(-\frac{R_{r=2}}{8}+\frac{1}{32}\right) \sin^4\theta +\frac{1}{16}\sin^2\theta,
\end{\eqn}
and the toroidal field $B_3$ is
\begin{\eqn}\label{B3}
\sin^2\theta B_3 = \frac{1}{r^2-2r} \bigg[ & &
-\frac{2\sin^2\theta}{r}\omega_3
              +\left(\frac{1}{r^2}-\frac{1}{4r}\right)\sin\theta\left(A_{2,\theta}-\frac{\sin\theta\cos^2\theta}{r^2}\right) \nn\\
& & +\frac{\sin^2\theta}{r^2}\left(\frac{1}{8}+\frac{1}{2r}\right)
-\left(\frac{\cos\theta}{4}A_2+i_3\right) \bigg],
\end{\eqn}
where we have made use of Eq.(\ref{notation}) and (\ref{2nd}). The
GS equation (\ref{grad-shafranov}) of the order of $O(a^4)$ is of
the following form,
\begin{\eqn}
\mathcal L A_4
&=&\omega_1 \left[\sin\theta\left(\frac{r^2}{8} -\frac{2}{r}\right)A_{2,r}\right]_{,r}
-\omega_3\left[-\left(1+\frac{2}{r}\right)\frac{1}{8}\sin^2\theta+2r\sin^2\theta B_1 \right]_{,\theta}\\
& &-\omega_1\left[-\sin\theta\left(1+\frac{2}{r}\right)\left(\frac{1}{8} A_{2,\theta}+\omega_3 \sin\theta\right)
   +\frac{\sin^2\theta\cos^2\theta}{4r^3}+2r\sin^2\theta B_3 \right]_{,\theta}\nn\\
& &+r^2\sin\theta B_1 \left[-\frac{1}{4}A_2+i_3'(A_0)\right] + \frac{1}{4}\cos\theta\sin\theta\left(\cos^2\theta B_1 + r^2 B_3\right)\nn\\
& &- \left(\frac{\sin\theta}{4r} A_{2,r}+\frac{2\cos^2\theta}{r^3\sin\theta}A_{2,r}\right)_{,r}
   +\left(\sin^2\theta B_3 + \frac{\cos^2\theta}{r^4\sin\theta}A_{2,\theta}-\frac{\cos^4\theta}{r^6}\right)_{,\theta}.\nn
\end{\eqn}
The convergence condition requires all source terms of the order
$O(1)$ should vanish, i.e.,
\begin{\eq}
0=\omega_3\left(\frac{1}{8}\sin^2\theta\right)_{,\theta}+\omega_1(\sin^2\theta\omega_3)_{,\theta}+r^2\sin\theta
B_1 i_3'(A_0)
    + \frac{1}{4}\sin\theta\cos\theta r^2 B_3 \ .
\end{\eq}
The above equation could be further simplified as
\begin{\eqn}
\omega_3(\sin^2\theta)_{,\theta}+ (\sin^2\theta\omega_3)_{,\theta}= i_{3,\theta} +2\frac{\cos\theta}{\sin\theta}i_3
=\frac{1}{\sin^2\theta}(\sin^2\theta i_3)_{,\theta} \Leftrightarrow \sin^2\theta \omega_3 = i_3,
\end{\eqn}
where we have used the result of Eq.(\ref{2nd}) and (\ref{B3}).
Together with  Eq.(\ref{h3w3}), we have that
\begin{\eqn}
i_3 &=& \frac{1}{2}\left(-\frac{R_{r=2}}{8}+\frac{1}{32}\right) \sin^4\theta +\frac{1}{32}\sin^2\theta,\nn\\
\omega_3&=& \frac{1}{2}\left(-\frac{R_{r=2}}{8}+\frac{1}{32}\right)\sin^2\theta + \frac{1}{32}>\frac{1}{32}.
\end{\eqn}
With the help of Eq.(\ref{notation}) and (\ref{2nd}), we finally
arrive at
\begin{\eqn}\label{omegaH}
\Omega=\Omega(A_\phi)&=&\frac{a}{8}+ a^3\omega_3,\nn\\
\sqrt{-g}F^{\theta r}=I(A_\phi)&=& \frac{a}{8}\sin^2\theta
+ a^3\left( \frac{1}{4}R(r)\sin^2\theta\cos^2\theta + i_3\right).
\end{\eqn}

\section{Discussion and Summary}
\subsection{Discussion}

\begin{figure*}
\includegraphics[scale=0.4]{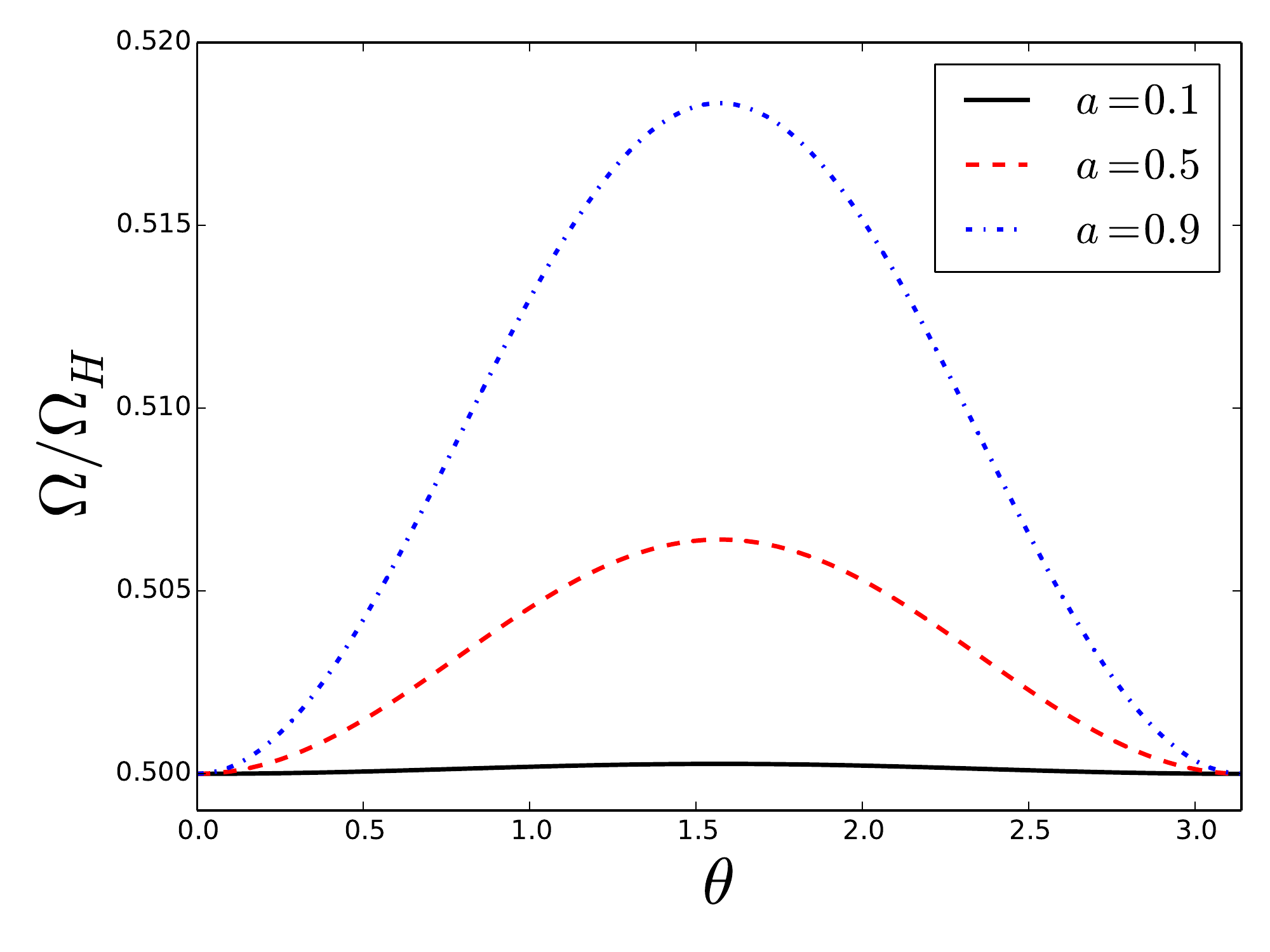}%
\includegraphics[scale=0.4]{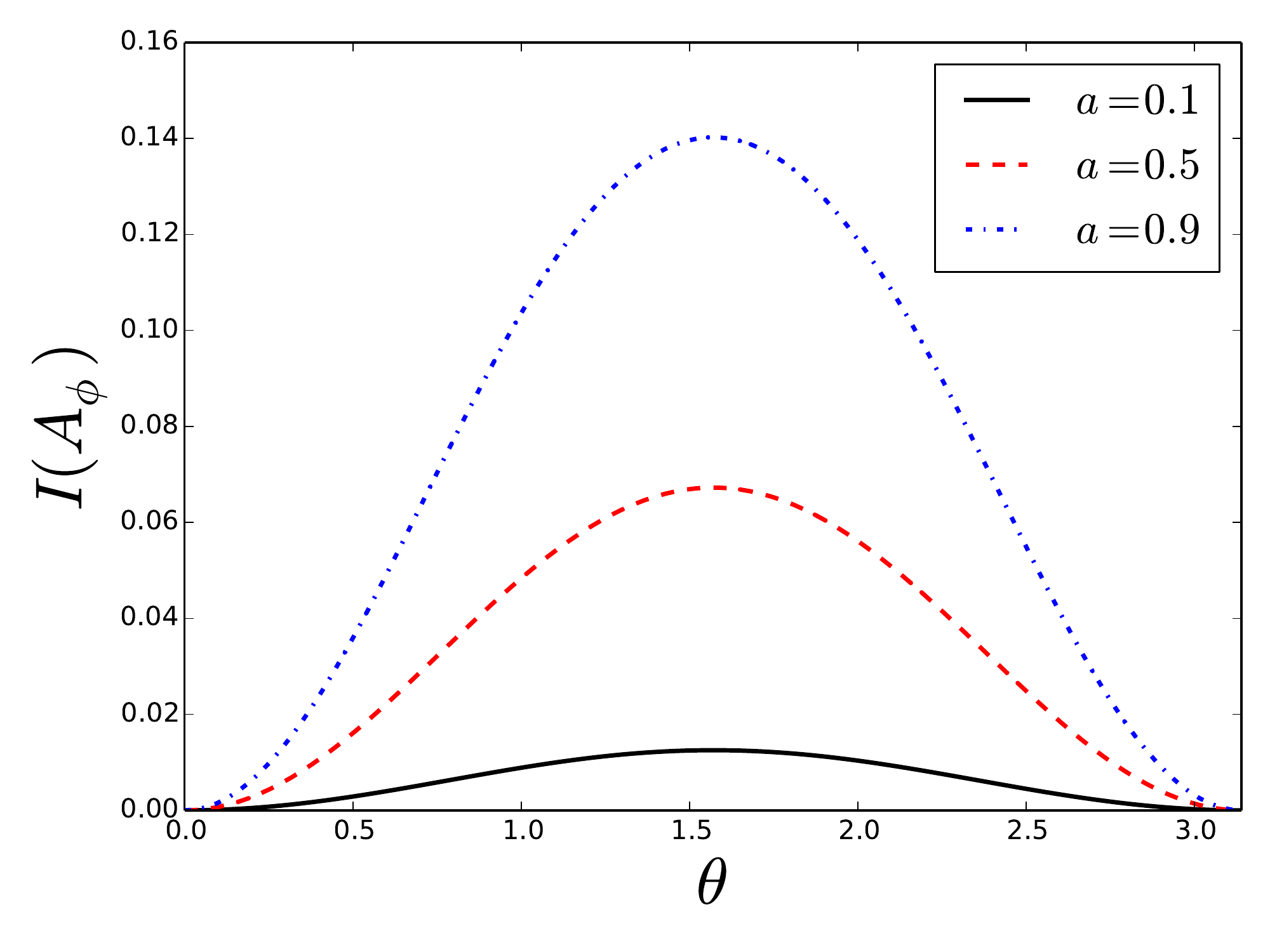}
\caption{Angular distribution of the ratio $\Omega/\Omega_H$ and the electric current $I$ on horizon ($r=2$),
where we keep the angular velocity of the BH accurate to fourth order, i.e., $\Omega_H=a/4 + a^3/16$ .}
\label{fig:omg}
\end{figure*}

The angular distribution of the fourth-order angular velocity
$\Omega$ and poloidal eletric current $I$ on the horizon is shown
in Fig.\ref{fig:omg}. For comparison, the corresponding simulation
results are also available (cf., Fig.1 and 2 of
\cite{Komissarov2001}). Both simulations and our analytic solution
imply that $\Omega = \Omega_H/2$ is a rather good approximation
for a wide range of black hole spins (at least for $a \lesssim
0.9$), where $\Omega_H = a/{2(1+\sqrt{1-a^2})}$ is the angular
velocity of the central BH. The fourth-order poloidal electric
current $I$ also shows better agreement with the simulation result
than the second-order one, especially for large spins.

The energy extraction rate, which is defined as $\dot E = -2\pi
\int_0^\pi \sqrt{-g} T^r_{\ t} d\theta = 2 \pi \int I(A_\phi)
\Omega(A_\phi) dA_\phi$ (\cite{Blandford1977d, Pan2014,
Beskin2009MHD}), could be written as
\begin{\eqn}
\dot E
&=& 2\pi a^2 \int  i_1 \omega_1 dA_0 + 2\pi a^4 \int i_1\omega_3 + \omega_1 i_3 + (i_1\omega_1 A_2)' dA_0\nn\\
&=& 2\pi a^2 \int  i_1 \omega_1 dA_0 + 2\pi a^4 \int \frac{1}{8}(i_3+\sin^2\theta\omega_3) dA_0\nn\\
&=& \frac{\pi}{24}a^2 + \frac{(56-3\pi^2)\pi}{1080} a^4
\end{\eqn}
where the prime denotes derivative with respect to $A_0$. Note
that the second term on the right hand side only depends on the
combination, $i_3+\sin^2\theta\omega_3$, which can be specified by
the Znajek horizon condition (i.e., Eq.(\ref{h3w3})). In fact,
this coincidence explains why \citet{Tanabe2008} could obtain the
correct energy extraction rate $\dot E$ without explicitly solving
$\Omega$ and $I$.

The stability is another interesting issue. Some analytic works
(\cite{Li2000,Lyutikov2006e,Giannios2006}) implied that the screw
instability may occur in the monopole perturbation solution due to
the Kruskal-Shafranov criterion. But no instability was noticed in
time-dependent GRMD (e.g. \cite{Komissarov2001,Komissarov2004e})
or GRMHD simulations (e.g. \cite{McKinney2009a,McKinney2013}). To
understand the discrepancy between analytic and numerical works,
\citet{Narayan2009g} (and \cite{Tomimatsu2001,McKinney2009a})
pointed out that Kruskal-Shafranov criterion may not be
appropriate for jet stability analysis, since it neglects the
stabilizing effect of the rotation of magnetic field lines.
According to the analysis of \citet{Tomimatsu2001}, which takes
the field rotation into account, the monopole perturbation
solution is possibly unstable only when $\Omega < \Omega_H/2$. Our
fourth order solution (i.e., Eq.(\ref{omegaH})) means that
\begin{\eq}
\Omega> \frac{1}{2}\Omega_H= \frac{a}{8} + \frac{a^3}{32} \ .
\end{\eq}
Obviously, the fourth-order monopole perturbation solution is
stable and is consistent with numerical simulations.

\subsection{Summary}
Two major difficulties are encountered in solving the GS equation
(\ref{grad-shafranov}): 1) it is a highly nonlinear second order
partial differential equation; 2) two proper boundary conditions
are necessary to uniquely specify the solution. The nonlinearity
could be partially removed by the perturbation technique. To fix
the boundary conditions problem, we impose the regularity
condition on the horizon (Eq.(\ref{znajek})) and the convergence
constraint (Eq.(\ref{convergence})). The latter one actually
serves as the boundary condition at infinity. With these two
boundary conditions, we re-establish the split monopole solution
to the order of $O(a^2)$ and get the new perturbation solution up
to the order of $O(a^4)$. By taking account of the stabilizing
effect of field rotation, we prove that the fourth-order monopole
perturbation solution is stable against the screw instability.

\begin{acknowledgments}
CY thanks the support by the National Natural Science Foundation
of China (grants 11173057 and 11373064), Yunnan Natural Science
Foundation (Grant 2012FB187, 2014HB048). Part of the computation
was performed at HPC Center, Yunnan Observatories, CAS, China.
\end{acknowledgments}

\bibliography{ms}

\end{document}